\documentclass[11pt,a4paper]{article}

\usepackage{jcappub}

\allowdisplaybreaks

\usepackage{ifthen}
\usepackage{graphicx}
\usepackage{dcolumn}
\usepackage{bm}
\usepackage{color}
\usepackage{amsmath}
\usepackage{amssymb}
\usepackage{subfigure}

\newcommand{\mpc}{\, {\rm Mpc}}

\newcommand{\lya}{Ly$\alpha$\ }

\newcommand{\vxperp}{\mathbf{x_\perp}}
\newcommand{\vkperp}{\mathbf{k_\perp}}

\newcommand{\vk}{\mathbf{k}}

\begin{document}

\title{Generating mock data sets for large-scale Lyman-$\alpha$ forest 
correlation measurements}

\author[a]{Andreu Font-Ribera,}
\author[b,c]{Patrick McDonald,}
\author[d,e]{Jordi Miralda-Escud\'{e}}

\affiliation[a]{Institut de Ci\`{e}ncies de l'Espai (IEEC/CSIC), 
Bellaterra, Catalonia}
\affiliation[b]{Lawrence Berkeley National Laboratory, California, US}
\affiliation[c]{Brookhaven National Laboratory, Long Island, US}
\affiliation[d]{Instituci\'o Catalana de Recerca i Estudis Avan\c cats, 
Catalonia}
\affiliation[e]{Institut de Ci\`{e}ncies del Cosmos (IEEC/UB), 
Barcelona, Catalonia}

\emailAdd{font@ieec.uab.es}

\abstract{ 
 Massive spectroscopic surveys of high-redshift quasars yield
large numbers of correlated \lya absorption spectra that can be used to
measure large-scale structure. Simulations of these surveys are
required to accurately interpret the measurements of correlations and
correct for systematic errors. An efficient method to generate mock
realizations of \lya forest surveys is presented which generates a field
over the lines of sight
to the survey sources only, instead of having to generate it over the
entire three-dimensional volume of the survey. The method can be
calibrated to reproduce the power spectrum and one-point distribution
function of the transmitted flux fraction, as well as the redshift
evolution of these quantities, and is easily used for modeling any
survey systematic effects. We present an example of how these mock
surveys are applied to predict the measurement errors in a survey with
similar parameters as the BOSS quasar survey in SDSS-III.
}


\maketitle

\section{Introduction}

  The hydrogen \lya absorption spectra of high-redshift sources are
being revealed as an extremely powerful tool for the study of
large-scale structure in observational cosmology. The numerous
absorption features observed in the spectra of quasars usually described
as the `` \lya forest '' were originally interpreted as discrete gas clouds,
but have been better understood and described as arising from the
continuous cosmic web of filamentary structures that is expected in the
Cold Dark Matter model of structure formation. Results from hydrodynamic
cosmological simulations have shown that the observed properties of the
\lya forest are generally in good agreement with the hypothesis of a
photoionized intergalactic medium with density fluctuations that are
related to the same primordial perturbations that give rise to the
galaxy distribution and the Cosmic Microwave Background fluctuations
(e.g., \cite{1998ARAA..36..267R,2006ApJS..163...80M}). The \lya forest 
spectra should
therefore be considered as a continuous field of the \lya transmitted
fraction $F({\bf x})$ (where ${\bf x}$ is the redshift-space coordinate),
which is related to the variations of the gas density, peculiar velocity
and temperature along the line of sight, and eventually to the primordial
density field, particularly on large scales, in which the complexities
of non-linear evolution become less important.

  In fact, if we have a large number of absorption spectra from
different sources covering a large volume and with a sufficiently dense
sampling, one can measure the redshift space power spectrum of the
field $F({\bf x})$. In the limit of large scales, this power spectrum
should be related to the linear power spectrum of density perturbations
as (see \cite{1999ApJ...520....1C,2000ApJ...543....1M,2003ApJ...585...34M})
\begin{equation}
 P_F(k,\mu_k)=b_{\delta}^2 (1+\beta \mu_k^2)^2\, P_L(k) ~,
\end{equation}
where $\mu_k$ is the cosine of the angle of the wavevector ${\bf k}$
in Fourier space relative to the line of sight, and $P_L$ is the
linear power spectrum of the mass density perturbations. This is the
same form of the linear power spectrum derived by \cite{1987MNRAS.227....1K} 
for any class of observed objects with a bias factor $b_{\delta}$, which
relates the amplitude of observed fluctuations to the amplitude of the
underlying mass fluctuations. But for the \lya forest, the redshift
distortion parameter $\beta$ depends on a second bias factor that is
related to the response of the mean value of $F$ to a large-scale
peculiar velocity gradient, and must be determined independently.

  Therefore, the promise of massive spectroscopic surveys of \lya
absorption spectra is to help determine the shape of $P_L(k)$ over a
wide range of scales and redshifts, and to use this to obtain crucial
cosmological measurements, such as the angular and redshift scale of the
Baryon Acoustic Oscillations, or the effect of neutrinos on the power
spectrum (e.g., \cite{2007PhRvD..76f3009M}). In addition, one can
determine the values of $b_{\delta}$ and $\beta$ at each redshift, which
are in principle predictable with hydrodynamic simulations from the
small-scale physics that determine the properties of the \lya forest
(\cite{2003ApJ...585...34M}). A first step in this direction was recently
accomplished by \cite{XiPush} from the first analysis of the
quasar absorption spectra in the BOSS survey.

  Accurately measuring the power spectrum requires a careful evaluation
and correction of any systematic errors that may be present in this
measurement in the analysis of real data. The only way to reliably doing
this is by generating several random realizations of the multiple \lya
absorption spectra in a survey, and introducing into them any
possible systematic effects to see how they may impact the inferred
power spectrum in the end. Some of the systematic effects that need to
be considered are the following: errors in the modeling of the quasar
continuum $C(\lambda)$, which is needed to evaluate the transmitted
fraction from the observed flux, $f(\lambda)=C(\lambda) F(\lambda)$;
variable spectral resolution and noise; flux calibration errors; the
impact of the redshift evolution of the \lya forest; the presence of
damped \lya , Lyman limit systems and metal absorption lines in the
spectra; or variations in the intensity of the cosmic ionizing
background. Modeling these systematic effects as
accurately and reliably as possible requires our ability to generate
mock surveys of \lya absorption spectra in large numbers, for many
different cases, and in a way that can be easily used. These mock
surveys must include a large number of lines-of-sight over large volumes
(like the ongoing BOSS survey in SDSS-III; \cite{2011arXiv1101.1529E}),
and somehow include the small-scale fluctuations of the \lya forest
that are present in the observed spectra of sources that are
point-like for practical purposes.

  Generating these mock surveys directly from three-dimensional
simulations, by selecting lines of sight from them, presents several
difficult challenges. The first is that having a large enough
volume to correctly simulate the power spectrum, at least up to scales
as large as the BAO peak, implies that the resolution of the simulations
cannot capture the smallest relevant scales for the \lya forest. In
addition, when using large three-dimensional simulations, the computer
resources that are required may not allow obtaining many mocks that are
independent, or changing the parameters of these mocks in an efficient
and fast way to enable a large number of tests.

  This paper presents a method to efficiently create these mock surveys
of \lya absorption spectra, taking advantage of the fact that the
transmitted fraction $F$ needs to be generated only on the discrete
lines of sight to the survey sources. The method consists of generating
one-dimensional fields for each line of sight and introducing
correlations among them as if they had been drawn from a
three-dimensional field. The capacity that is lost with this method is
using hydrodynamic simulations that include the non-linear gravitational
evolution of density fluctuations and other physical effects to simulate
the field $F({\bf x})$. However, if we care only about the large-scale
power spectrum of this field and the errors to which it can
be measured, it is in principle enough to ensure that the mocks have
the same variance in the small-scale fluctuations to reproduce their
effect on large scales. The way the mock surveys are generated is by
using an input power spectrum of $F({\bf x})$ in redshift space that
includes a non-linear correction for small scales, and which is assumed
to be calibrated from the results of cosmological simulations with
enough resolution or directly from the observational results. The mocks
can also incorporate any desired one-point distribution of $F$ and
the redshift evolution of both the power spectrum and the distribution
of $F$.

  Hence, the philosophy of these mock surveys is that they are
generated from an input model of the power spectrum and other
quantities, and that they should be used for predicting the large-scale
correlation measurements of the \lya forest and the way they are
affected by any systematic errors that can be introduced. However, the
field $F({\bf x})$ that is simulated is purely local and inferred from
the linear overdensity, so it does not reproduce the 3-point or higher
n-point correlations of the \lya forest.

  The method is presented in detail in \S 2, and an application to an
example of a survey similar to BOSS is presented in \S 3. Another
application of these mocks to simulate the effect of damped \lya
systems is discussed in \cite{LyaDLA}. This method was
already used for simulating the sample of spectra used in 
\cite{XiPush}, and is being improved for application to the final BOSS survey.

A standard flat $\Lambda CDM$ cosmology is used in this paper with the
following parameters:
$h=0.72$ , $\Omega_m=0.281$, $\sigma_8=0.85$, $n_s=0.963$, $\Omega_b=0.0462$.

\section{Method to generate mocks of correlated \lya spectra}
\label{sec_method}

  A \lya forest spectrum is given by the fraction of transmitted flux,
$F=\exp(-\tau)$, where $\tau$ is the optical depth, at each observed
wavelength. We define the comoving coordinate in redshift space, $x$,
related to the wavelength by $dx = c/H(z) (d\lambda /
\lambda_{\alpha})$, where $H(z)$ is the Hubble constant, the redshift
is $1+z=\lambda/\lambda_{\alpha}$ and $\lambda_\alpha=1216\, {\rm \AA}$
is the \lya resonance wavelength. The observed spectrum is the
product of $F(x)$ times the continuum of the source, which is not
independently observed and must be modeled. We shall not deal in this
paper with the issue of modeling the continuum. Our mocks are
realizations of the function $F(x)$ on multiple, correlated lines of
sight.

  In this paper we shall generally work with the variable
\begin{equation}
 \delta_F(x) = {F(x) \over \bar F} - 1 ~,
\label{dFdef}
\end{equation}
where $\bar F$ is the mean value of $F$ at a given redshift.
All the 2-point correlations
appearing in this article are of this $\delta_F$ variable unless 
otherwise stated. This section describes the method to generate a set
of mock \lya spectra with any specified distribution function and power
spectrum for the $\delta_F$ variable. The main idea for the case of a
Gaussian field is explained in \ref{ss_gauss}, which is then generalized
to any desired distribution of $\delta_F$ (\ref{ss_flux}). The inclusion
of redshift evolution is discussed in \ref{ss_evol}.

\subsection{Generation of a Gaussian random field}
\label{ss_gauss}

  The most important requirement that our mock \lya spectra must meet
if they are to accurately predict any systematic and statistical errors
in the measurements of large-scale correlations in $\delta_F$ is that
they have a redshift space power spectrum of the flux that accurately
matches the observed one. In this way, the intrinsic variance of the
\lya absorption at any scale can be reproduced, and the way it affects
the sampling errors on all other scales is correctly taken into account.
Our method to generate mock \lya spectra can take as input any desired
power spectrum $P_F(k_{\parallel}, k_{\perp})$ in redshift space, where
$k_{\parallel}$, $k_{\perp}$ are the components of the wave vector in Fourier
space parallel and perpendicular to the direction of the line of sight.

\subsubsection{Sampling the volume unevenly}

  The usual way to generate a Gaussian random field in realizations of
cosmological perturbations is to generate first a set of independent
Fourier modes in a three-dimensional cubic box with a specified power
spectrum, and then doing the Fourier transform to obtain the real-space
field. This method yields the value of the field at all the cells in the
cubic volume at once.

  However, to simulate the measurement of correlations up to the BAO
scale in a survey of quasar spectra, we need to cover a volume with a
size of at least several times the BAO scale, with a required resolution
needed to capture the fluctuations in the low-density intergalactic
medium of at least $\lambda_J/(2\pi)= \sqrt{3/2}\, c_s t$, or $\sim 100$
comoving kpc (where $\lambda_J$ is the Jeans length, $c_s$ is the sound
speed of the intergalactic gas, and $t$ the age of the universe; 
well-resolved simulations of the IGM typically use cells a factor of a few 
smaller than the Jeans length; see, e.g., \cite{2005ApJ...635..761M,
1980lssu.book.....P}).
The minimum dynamic range from the smallest to the largest scale is then 
$\sim 10^4$, or $10^{12}$
simulated points (and even larger if the entire volume of a survey like
BOSS is to be generated), which results in a serious computational
problem for being able to easily generate large numbers of mocks in a
simple way.

  Our method uses the fact that we are only interested in the values of
the field along a number of infinitely thin lines of sight traced by the
quasar light. Hence, we can generate a Gaussian field on these
one-dimensional lines only, and introduce correlations among them
directly in real space. A first, simple-minded way to achieve this might
be to first generate an independent Gaussian variable at each pixel, $g_i$,
and then combine them to generate the final field $\delta_{gj} = L_{ij} g_i$
which has the desired correlation $C_{ij}$:
\begin{equation}
C_{ij} = \, < \delta_{gi} \delta_{gj} > \, = \, < L_{ik} g_k L_{jl} g_l > \, =
 L_{ik} L_{jl} \delta_{kl} = L_{ik} L_{jk} ~ .
\end{equation}
A particularly efficient way to obtain the required matrix $L$ for the 
transformation is the result of the Cholesky decomposition of the covariance 
matrix $C$, i.e., a lower triangular matrix $L$ obeying $C = L L^T$. 
Numerically, there are several algebraic packages that perform the Cholesky 
decomposition very efficiently.

  For a practical application, the number of pixels that are needed to
model a typical observed spectrum and to include the power down to the
smallest relevant scales is $N_p \sim 10^3$ for each line of sight. For
a survey with $N_q$ quasars, the total number of elements of the
correlation matrix C that need to be computed is $(N_p \times N_q)^2$.
Clearly, this method would break down for a relatively small number of
quasars. Fortunately, there is a better way to do it.

\subsubsection{Parallel lines of sight}

  Let us assume for the moment that the lines of sight in the survey
are perfectly parallel. Let $\delta_g(x_{\parallel},\vxperp)$ be the
correlated Gaussian variable we want to generate at the position
$x_{\parallel}$ of the line of sight at coordinate $\vxperp$. We can do
the one-dimensional Fourier transform of $\delta_g$ on the direction of
the line of sight only, to obtain $\tilde\delta_g(k_\parallel,\vxperp)$.
These one-dimensional Fourier modes have the following correlation:
\begin{align}
\label{eq:f1dc}
\left<\tilde\delta_g\left(k_\parallel,\vxperp\right) 
      \tilde\delta_g\left(k_\parallel^\prime,\vxperp^\prime\right)\right>
 &=\frac{1}{2\pi} \int d\vkperp \exp(i\vkperp\vxperp) 
                   \int d\vkperp^\prime \exp(i\vkperp^\prime\vxperp^\prime) \\
\nonumber &\times \delta^D\left(k_\parallel+k_\parallel^\prime\right) 
    \delta^D\left(\vkperp+\vkperp^\prime\right)P\left(\vk\right) \\
\nonumber &=2\pi\delta^D\left(k_\parallel+k_\parallel^\prime\right)
    P_\times\left(k_\parallel,\left|\vxperp-\vxperp^\prime\right|\right) ~,
\end{align}
where the symbol $\delta^D$ stands for the Dirac delta function,
$P(\vk)$ is the power spectrum of $\delta_g$, and
\begin{equation}
P_\times\left(k_\parallel,r_\perp\right) =
\frac{1}{2 \pi} \int_{k_\parallel}^\infty k~dk ~
J_0\left(k_\perp r_\perp\right) ~P\left(k_\parallel, k_\perp \right) ~.
\label{eq:cp}
\end{equation}
The crucial property is that the one-dimensional modes $\tilde\delta_g$
on different lines of sight are independent except when $k_\parallel =
k_\parallel^\prime$. Therefore, the problem is now separated for each
value of $k_\parallel$, and the Cholesky decomposition operation needs
to be performed on $N_p$ matrices of size $N_q \times N_q$ only.

  Hence, the procedure to be followed in our method is as follows. We
first choose a grid of values of $k_\parallel$ for the Fourier
transforms on the line of sight. 
For each value of $k_\parallel$, we compute the correlation of the 
one-dimensional Fourier modes for every pair of lines of sight, using 
equations (\ref{eq:f1dc}) and (\ref{eq:cp}). Each one of these 
$N_q \times N_q$ matrices, $C_k = P_\times(k_\parallel,r_\perp)$, is
then Cholesky-decomposed to obtain a matrix $L_k$. After generating a
set of independent Gaussian variables for each quasar and each value of
$k_\parallel$, $g_{kq}$, we compute the new set $\tilde\delta_g =
L_k g$, and we then do the inverse one-dimensional Fourier transform
of these to finally obtain the $\delta_g$ variables, with all the real
space correlations that are implied by the input 3-d power spectrum
$P(\vk)$.

  In reality, the \lya spectra need to be generated for quasars that
are at different redshifts. We do this by first generating the spectra
lines of sight of a long enough comoving length $L$, evaluating
$\delta_g$ on bins of comoving width $\Delta x$. We set the center of
the line of sight at a central redshift $z_c$ (we use $z_c =2.6$ in this
paper), and every bin is then mapped into a redshift according to its
comoving coordinate. We then use only the part of the spectrum of each
quasar that is in the restframe wavelength range for \lya forest
analyses. We use 1041 \AA $ < \lambda_r < 1185$ \AA \, in this paper,
the usual range to avoid Ly$\beta$ contamination and the proximity
effect zone near the quasar. We also use $L=4096\, h^{-1}$ Mpc, long
enough to make any periodicity effects negligible, and 
$\Delta x=0.5 h^{-1}$ Mpc, slightly smaller than the typical pixel width
in the BOSS spectrograph ($1\, {\rm \AA} \simeq 0.7\, h^{-1}$ Mpc at the
redshifts of interest).

\subsection{Flux distribution}
\label{ss_flux}

  The principal goal of the mocks of correlated \lya forest spectra we
want to generate is to simulate the observed spectra in a survey like
BOSS that includes all of the statistical and systematic errors we may
consider to obtain a correction for them when computing any statistical
property. It is therefore important that the perturbation in the
transmitted flux fraction, $\delta_F$, in the mock spectra has the same
distribution as the observed one,
in order that the impact of continuum fitting and noise on the measured
correlations and their errorbars are correctly simulated. Note that the
value of the noise that is added in the mocks and the way that the
continuum fitting is obtained will depend on a complex way on the values
of $\delta_F$. Here we
generalize our method to generate a field $\delta_F$ with the desired
probability distribution function $p_F(\delta_F)$ and any power spectrum
$P_F(\vk)$. Although the higher order n-point correlations of $F$ will
obviously still be different for the mocks and the real \lya forest
spectra, we expect this to have negligible impact on the computed errors of any
statistical measurements on large scales
(e.g., \cite{2006ApJS..163...80M} found that errors computed as if the flux 
field was Gaussian were close to errors determined by bootstrapping over real
spectra, even on Mpc scales). 

  This generalized method consists of generating first our field
$\delta_g$ 
with a Gaussian distribution,
$p_g(\delta_g)=\exp(-\delta_g^2/2)/\sqrt{2\pi}$, with a different power
spectrum $P_g$ such that, after transforming the field to the new
variable $\delta_F(\delta_g)$, the desired probability distribution
function $p_F(\delta_F)$ and power spectrum $P_F$ are obtained.
The required transformation $\delta_F(\delta_g)$ is obtained by
integration of the equation
\begin{equation}
 \frac{d\delta_F}{d\delta_g} = \frac{p_g(\delta_g)}{p_F(\delta_F)} ~.
\end{equation}

  Let us consider the correlation functions $\xi_F(r_{12})$ and
$\xi_g(r_{12})$ of the field values at two points ${\bf x}_1$ and
${\bf x}_2$ separated by the distance $r_{12}$. We designate these
field values as $\delta_{F1}$, $\delta_{F2}$, $\delta_{g1}$,
$\delta_{g2}$. Since the field $\delta_g$ is strictly Gaussian, the
correlation functions are related by
\begin{align}
\label{eq_xi_gauss}
  \xi_F(r_{12}) = & \left< \delta_{F1}\, \delta_{F2} \right> \\
 \nonumber    = & \int_{-1}^{1/\bar F-1} d\delta_{F1} 
 \int_{-1}^{1/\bar F -1} d\delta_{F2}\, p_{2F}(\delta_{F1},\delta_{F2})
 \,  \delta_{F1} \delta_{F2} \\
 \nonumber    = & \int_{-\infty}^\infty d\delta_{g1}
    \int_{-\infty}^\infty d\delta_{g2}\, p_{2g}(\delta_{g1},\delta_{g2})
 \,  \delta_{F1} \delta_{F2} \\
 \nonumber    = & \int_{-\infty}^\infty d\delta_{g1}
                  \int_{-\infty}^\infty d\delta_{g2}\,
                  \dfrac{ \exp{ \left[-\dfrac{\delta_{g1}^2 + \delta_{g2}^2
                        -2\delta_{g1}\, \delta_{g2}\, \xi_g(r_{12})}
                             {2(1-\xi^2_g(r_{12}))} \right] } }
                       {2 \pi \sqrt{1-\xi^2_g(r_{12})}} \,
                  \delta_F(\delta_{g1})\, \delta_F(\delta_{g2}) ~.
\end{align}

Note that we have assumed that the Gaussian field has unit variance, without
loss of generality as an overall normalization factor can always be included
in the definition of the mapping $\delta_F(\delta_g)$.
This relation between the two correlations $\xi_F$ and $\xi_g$ is
actually a one-dimensional function that is totally independent of the
separation $r_{12}$ or any other variable: it depends only on the relation
$\delta_F(\delta_g)$. We can therefore tabulate and invert the
function $\xi_F(\xi_g)$.

  The procedure to generate a random field $\delta_F$ is therefore the
following: we start with an input model for the three-dimensional power
spectrum $P_F$ of the flux transmission, and compute the Fourier
transform to obtain $\xi_F$. We then convert this to the correlation
function $\xi_g$, and proceed to compute the correlations of
one-dimensional power for the Gaussian field $g$ in equation
\ref{eq:cp}), which can be re-expressed as:
\begin{equation}
  P_{g\times} (k_\parallel, r_\perp) = \int_{-\infty}^\infty
 dr_\parallel\, e^{i k_\parallel r_\parallel} \xi_g(r_\parallel, r_\perp) ~.
\end{equation}
We mention here that this procedure does not in general work for any
distribution function $p_F(\delta_F)$, because sometimes the resulting
power $P_{g\times}(k_\parallel,r_\perp=0)$ may be negative for some values of 
$k_\parallel$. An auto-power spectrum must be positive definite, because the
variance of Fourier modes can never be negative (for $r_\perp \neq 0$, 
$P_{g\times}(k_\parallel,r_\perp)$ can be negative). 
Fortunately, this does not occur for the input model chosen here, but
it may well occur with other distributions 
(see \cite{1992MNRAS.259..652W} for a discussion of the same problem in 
the context of non-Gaussian initial conditions).

\subsection{Redshift evolution and non-parallel lines of sight}
\label{ss_evol}

  The power spectrum of $\delta_F$ is a function of redshift. The main
evolution is in the amplitude of the power spectrum, but a more general
evolution in the shape is likely to be present, particularly on small
scales. To introduce the redshift evolution in our model, we generate
the field $\delta_F$ for several discrete values of the redshift,
obtaining a set of realizations $\delta_{Fi}(x_\parallel, x_\perp)$,
where the subindex $i$ labels the redshift. Each of these realizations
is generated with the same amplitudes and complex phases of the Fourier
modes $\tilde\delta_g$, and varying only the amplitude of the power
spectrum that is different due to the evolution with redshift. In other
words, the realizations at different redshifts have all the same random
elements and change only because of the variation in the power spectrum,
so we can smoothly interpolate between them to obtain the generated
field at any desired redshift.

  The effect of the variation of the angular diameter distance and
Hubble constant with redshift, and the fact that the lines of sight are
not parallel, is included in the same way as the redshift evolution. The
power spectrum can be expressed in terms of fixed angular and redshift
separations at the discrete values of the redshift at which the multiple
fields $\delta_{Fi}$ are generated.

  The final field $\delta_F$ is obtained by linear interpolation of the
multiple fields as the redshift varies along the lines of sight,
introducing in this way the gradual evolution in the power spectrum
amplitude, the Hubble constant and the angular diameter distance with redshift.

  In this paper, the redshift values at which the fields $\delta_{Fi}$
are generated are $z=1.96$, 2.44, 2.91, and 3.39.

\subsection{Input model for \lya forest mocks used in this paper}
\label{ss_modelp}

  The distribution and power spectrum of the transmitted flux fraction
can be determined from observations and can also be computed in theory
from hydrodynamic cosmological simulations of the intergalactic medium.
As observational progress is made, mocks of \lya forest surveys can be
adjusted to reproduce as accurately as possible the observational
determinations of the distribution and power spectrum of $\delta_F$,
which guarantees an accurate modeling of the measurement errors for any
quantities. Here, we use the parameterized fitting formula introduced by
\cite{2003ApJ...585...34M} to fit the results of the power spectrum 
from several numerical simulations,
\begin{equation}
 P_F(k,\mu_k) = b_{\delta}^2 (1+\beta \mu_k^2)^2 P_L(k) D_F(k,\mu_k) ~,
\label{ips}
\end{equation}
where $b_{\delta}$ is the density bias parameter at $z=2.25$,
$\beta$ is the
redshift distortion parameter, $\mu_k=k_{\parallel}/k$, $P_L(k)$ is the
linear matter power spectrum, and $D_F(k,\mu_k)$ is a non-linear term
that approaches unity at small $k$. This form of $P_F$ is the expected
one at small $k$ in linear theory, and provides a good fit to the
observations reported in \cite{XiPush}.
Note that we do not generate a density and a velocity field, but we
directly generate the \lya forest absorption field instead, with the
redshift distortions being directly introduced in the input power
spectrum model of equation (\ref{ips}), with the free parameter $\beta$
that measures the strength of the redshift distortion.

  We use the parameters given in the central model of 
\cite{2003ApJ...585...34M},
$b = -0.1315$ and $\beta=1.58$ (the negative sign of $b$ simply
reflects the decrease of $\delta_F$ with gas density, and does not
affect any equations in this paper because it always appears as $b^2$).
Only the amplitude of the 
power spectrum is assumed to evolve with redshift, following a power-law:
\begin{equation}
 P_F(k,\mu_k,z) = P_F(k, \mu_k, z=2.25)
 \left( \frac{1+z}{1+2.25} \right)^\alpha ~.
\end{equation}
We use the value $\alpha=3.8$ in this paper, as suggested by the evolution of 
the one-dimensional P(k) measured in \cite{2006ApJS..163...80M}.

  For the probability distribution, we use a log-normal model for the
optical depth $\tau$,
\begin{equation}
 \label{eq:lognormal}
 F = e^{-\tau} =  \exp{ \left( -a e^{\gamma g}\right)} ~,
\end{equation}
where $g$ is a Gaussian variable of unit dispersion, and $a$ and $\gamma$ are
two free parameters determining the mean transmission $\bar{F}$ and 
its variance.

In the future, a new distribution for $F$ that
more accurately matches the observed one should be used for the mocks,
but the log-normal approximation suffices for the purpose of this paper
of demonstrating the applications of \lya forest mocks.

  We assume a mean transmitted fraction that approximately matches the
observations (\cite{2006ApJS..163...80M}): 
\begin{equation}
 \ln \bar{F} (z) = \ln(0.8) \left( \frac{1+z}{3.25} \right)^{3.2} ~.
\label{meanf}
\end{equation}
The values of $a$ and $\gamma$ at each redshift can be derived by
requiring the mean value of $F$ to match equation (\ref{meanf}), and its
dispersion to reproduce the value implied by the power spectrum $P_F$.
The result for the parameters at the four redshifts we use are the
following: $ a = 0.065$ and $\gamma = 1.70$ at $z=1.96$;
$ a = 0.141$ and $\gamma = 1.53$ at $z=2.44$;
$ a = 0.275$ and $\gamma = 1.38$ at $z=2.91$; and
$ a = 0.487$ and $\gamma = 1.24$ at $z=3.39$.

\section{Results}
\label{sec_res}

  This section presents the results for the characteristic errors in the 
measurement of the correlation function, as an example of a simulated
\lya forest survey with similar characteristics as BOSS. 

\subsection{Model for the quasar survey}

The first step to generate a mock \lya forest survey is to generate the
quasar sample. We randomly distribute quasars (with no clustering) over
a circular area $A=300\, {\rm deg}^2$ and the redshift range $2.15 < z <
3.5$, following the quasar luminosity function measured in 
\cite{2006AJ....131.2788J} up to a limiting magnitude of $g=22$. 
We select only 75 \%
(independently of g magnitude and redshift) of the quasars in order to
have a quasar number density closer to the one obtained in the BOSS
survey ($\sim 15-17 \, {\rm deg}^{-2}$). 
The total number of quasars in the sample is $N_q \simeq 5000$. The code
we use to generate the absorption fields with the method described in
Section 2 was able to generate all the absorption spectra in one survey
mock with a node with 8 CPU in a few hours. 

The redshift distribution of the sources in a real survey usually
differs substantially from that inferred from the model luminosity
function, mainly because the target selection efficiency has a strong
dependence on redshift. In particular, in the optical color selection
used by SDSS, quasars at $z\sim 2.7$ overlap the stellar locus and
are confused with stars, making them harder to select.
There is also a change in efficiency as a function of the foreground
stellar density and dust absorption. We do not include these effects
here. If anything, these effects should reduce the errors of measuring
the \lya correlation because they should cause an increased overlap of
the \lya spectra redshift range and an increased number of quasar pairs
at small separations, for fixed mean quasar density.

  After having constructed the spectra of the transmitted fraction $F$
as described in the previous section, we generate a realistic observed
quasar spectrum that includes the spectral resolution and noise
approximately matching those in the BOSS survey, following these steps:
\begin{itemize}
 \item A new set of pixels for a mock of the physical spectrum in units
  of flux is constructed, covering the whole, fixed wavelength range
  $3600 $ \AA $\, < \lambda < 9000 $ \AA, with pixels of constant
  wavelength width $\Delta \lambda = 1$ \AA. The width of these pixels
  in comoving separation is therefore changing along the spectrum.

  For each quasar, we compute the mean value of the pixel width 
  ($R_w$) and the mean value of the Point Spread 
  Function($R_p$) in comoving separation, over the region
  that is used for measuring the \lya forest correlation function,
  1041 \AA $\, < \lambda <$ 1185 \AA\ in the rest frame. We then convolve
  our spectrum of $\delta_F$ in the original pixels of constant comoving
  length with the Point Spread Function that results from the convolution
  of a Gaussian spectral resolution and the pixel width in the final
  wavelength bins:
\begin{equation}
 \delta_F (x) = \frac{1}{2\pi}  
                \int dk \, e^{i k x} \tilde\delta_F(k) 
              ~ \exp \left[-\frac{k^2 R_p^2}{2} \right] 
              ~ \left[ \frac{\sin(k R_w / 2)}
                      {k R_w / 2} \right]^2 ~ . 
\end{equation} 
The values of $R_p$, $R_w$ depend on the quasar redshift, with 
typical values being in the range $0.6-0.8\, h^{-1}$ Mpc. We note that
the wavelength dependence of the spectral resolution and the pixel width 
in the BOSS spectrograph are actually quite complex, and they should be
carefully treated if one is interested in small-scale correlations. 

 \item Each pixel in the spectrum with constant wavelength bins is
  assigned the value of $F$ in the nearest bin of the spectrum with
  pixels of constant comoving width. We set $F=1$ for 
  wavelengths outside the \lya forest range.

%
%

\item We multiply the spectrum of $F$ by the continuum for each quasar,
using the mean rest-frame spectra obtained in \cite{2005ApJ...618..592S}. 
A spectrum of physical flux, $f(\lambda)$, is
obtained after normalizing to match the $g$ magnitude of the quasar.

 \item The expected noise variance for the case of the BOSS spectrograph
  with an exposure time of 1 hour is computed
  at each pixel using the expression

  \begin{equation}
   \label{eq:noise}   
   \sigma_N^2 (\lambda) 
         = A + B(\lambda) ~ \left[ f(\lambda) + s(\lambda) \right] 
                    ~ \Delta\lambda ~ ,
  \end{equation}
  where $s(\lambda)$ is a typical sky flux in BOSS, $A$ is the read-out 
  noise and $B(\lambda)$ is related to the BOSS throughput.
  These functions have been estimated using BOSS survey code
  \cite{2009astro2010S.314S}.

 \item We add a Gaussian random variable with variance $\sigma_N^2$ to the 
  flux $f(\lambda)$ at each pixel, and divide the resulting flux by the 
  continuum to obtain a new spectrum of transmitted fraction $F$ (which is
  no longer restricted to the range $0 < F < 1$ because of the noise that
  has been added).
\end{itemize}

The detailed properties of the noise in the real survey are more
complicated, but this simple procedure allows us to approximately study
the effect of noise on the correlation function measurement.

\begin{figure}[h!]
 \begin{center}
  \includegraphics[scale=0.7, angle=-90]{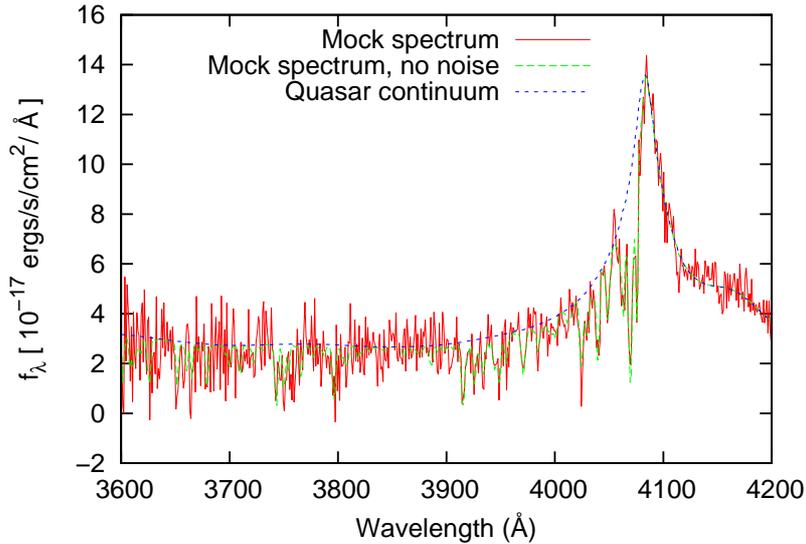}
 \end{center}
 \caption{Mock quasar spectrum ({\it solid red}), without noise 
({\it dashed green}) and without \lya absorption ({\it dashed blue}).}
 \label{fig:example}
\end{figure}

An example of mock spectra with continuum and noise added is shown in
Figure \ref{fig:example}. 

We have generated 50 realizations of this mock survey to obtain the 
results that are presented next.

\subsection{Measurement of the Correlation Function}

  We estimate the value of the correlation function as the weighted average 
of the product of the $\delta_F$ variable in all pixel pairs that have a
redshift space separation $r$, angle cosine $\mu$ and mean redshift $z$,
which are within a certain bin of width $\Delta r$, $\Delta \mu$ and
$\Delta z$, which we designate as $A$:
\begin{equation}
 \label{eq:estimator}
  \hat \xi_{A} = \frac{\sum_{i,j \in A}  w_i w_j \, 
             \delta_{Fi} \delta_{Fj}}{\sum_{i,j \in A} w_i w_j} ~.
\end{equation}
The weights $w_i$ for each pixel are to be chosen to minimize the error
of the correlation function estimator, although special care needs to
be taken on any possible bias that they may introduce.
Our first choice for the weights was to set them equal to
the total inverse variance in each pixel. The total variance is equal
to the sum of the intrinsic \lya forest fluctuations,
$\sigma^2_F(z) = \left< \delta_F^2 \right>$,
and the variance caused by the instrumental noise,
$\sigma^2_N(\lambda) / \left[ \bar{F}(z) C(\lambda) \right]^2$,
so the weights are
\begin{equation}
 \label{eq:weight}
  w_i = \sigma_i^{-2} = \left[ \sigma_F^2(z_i) 
     + \dfrac{\sigma_N^2(\lambda_i)}
             { \left\{ \bar{F}(z_i) ~ C(\lambda_i) \right\}^2} \right]^{-1} ~.
\end{equation}
These are the same weights that were adopted in \cite{XiPush}.
The effect of the intrinsic variance correlation in neighboring pixels
is ignored, as was done in
\cite{XiPush}, since this is not expected to have any effects on large
scales. Note that the noise variance in equation \ref{eq:noise} applies
to the flux variable $f(\lambda) = \bar{F} [1+\delta_F(\lambda)] \,
C(\lambda)$. The corresponding contribution to the variance of
$\delta_F$ in equation \ref{eq:weight} is obtained by dividing by
$\left[ \bar{F(z)} C(\lambda) \right]^2$.

\begin{figure}[h!]
 \begin{center}
  \includegraphics[scale=0.5, angle=-90]{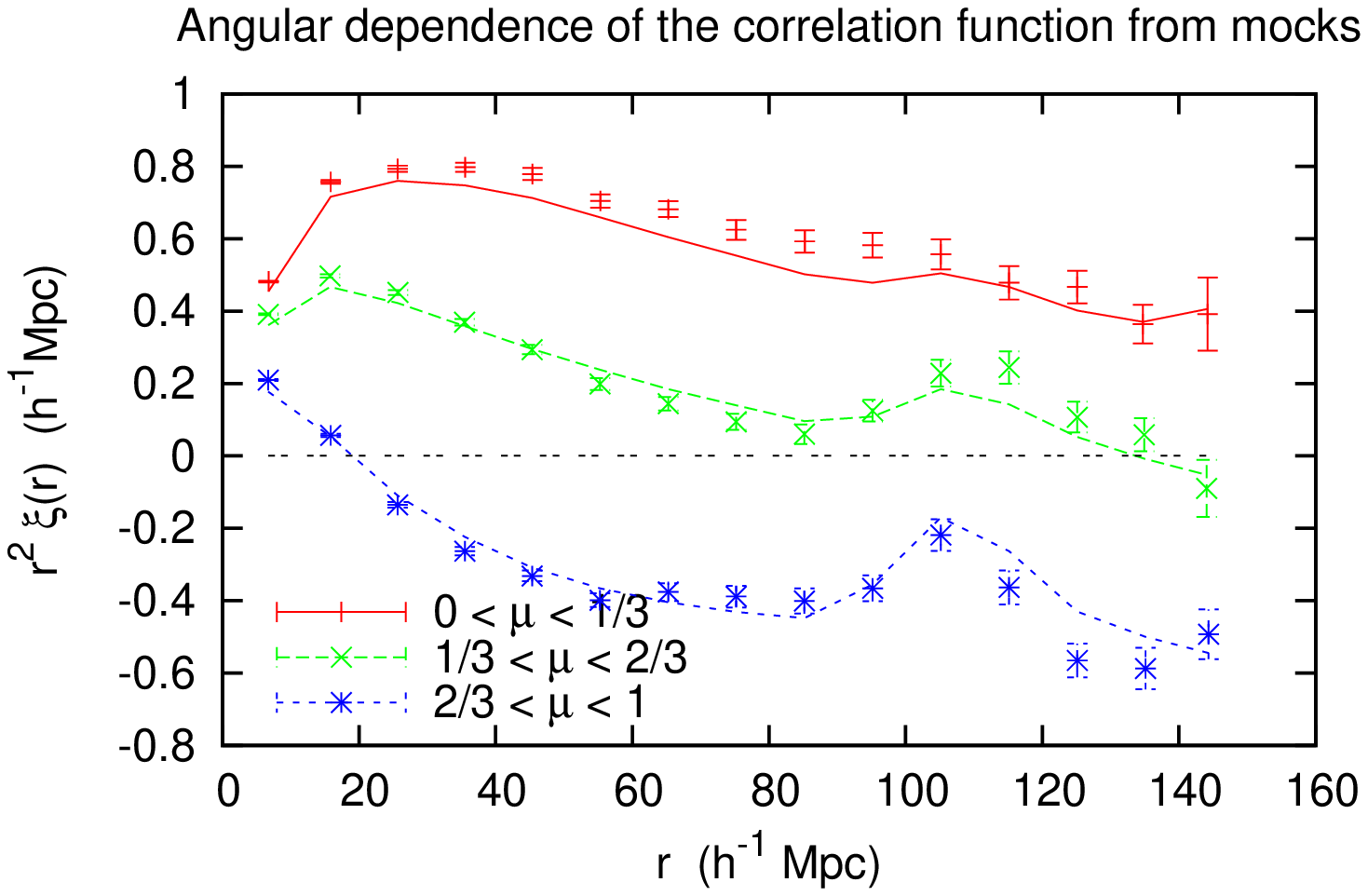}
  \includegraphics[scale=0.5, angle=-90]{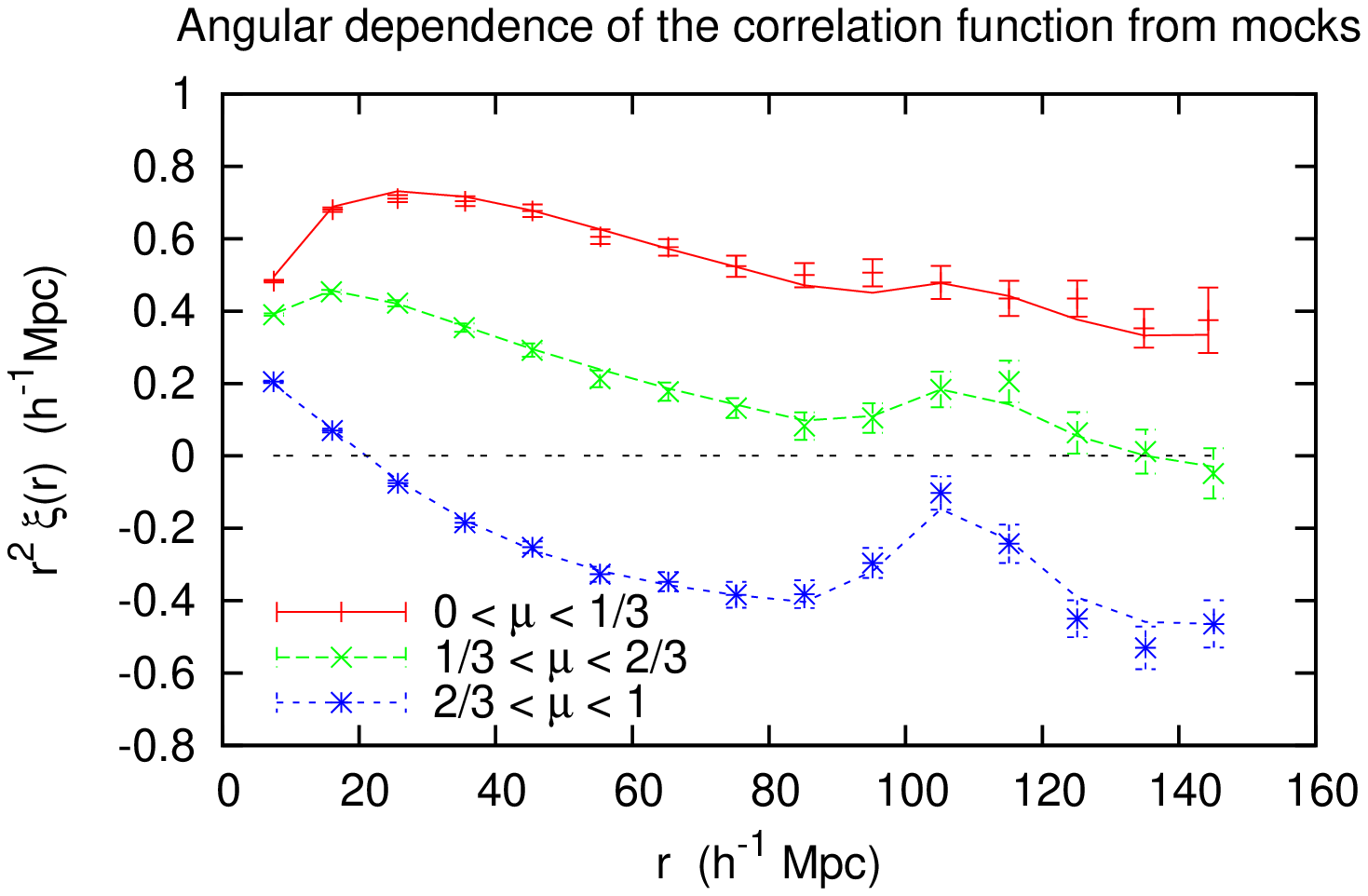}
 \end{center}
 \caption{Correlation function obtained from the average of 50 mocks of
 our survey model ({\it points}), as a function of $r$ in three different
 bins in $\mu=cos(\theta)$, compared to the input model ({\it lines}).
 The errorbars are computed for the average of 50 mocks, from the dispersion
 between them.
 {\it Left panel:} Pixel weights from equation \ref{eq:weight}, which 
introduce a bias owing to their dependence on the measured flux.
 {\it Right panel:} Pixel weights from equation \ref{eq:unbiased}, which 
are unbiased.}

 \label{fig:wedges}
\end{figure}

  The correlation function is estimated first on a large number of small
enough bins for the final result to have converged to the correct value,
using 150 bins in $r$ up to $r=150\, h^{-1}\, {\rm Mpc}$, 20 bins in
$\mu$ and 20 bins in $z$, all of them linearly spaced.
Then, for the purpose of plotting the results, the correlation function
is averaged over all redshifts and compressed
into broader bins of $10 \, {\rm Mpc}/h$ in $r$, and three bins
in $\mu$. This average over the small bins into the broader bins is done
using the combined weights of each of the small bins, in the same way
for the theoretical value of the correlation function and its estimate
obtained in each mock. The results are plotted in the left panel of
Figure \ref{fig:wedges}, where the average estimate from the 50 mocks
is shown as the points with errorbars, and the theoretical value from the
model power spectrum used to generate the mocks is shown as the curves.

\vspace{1cm}

  The errorbars reflect the expected errors for the average of 50 mocks,
and have been computed from the dispersion between the 50 mocks. Therefore,
the true errors of one of our realizations of a survey with an area of
$300$ square degrees are larger by a factor $\sqrt{50-1}=7$.
This neglects the edge
effects of the survey (the fact that fewer pairs of quasars are found
for quasars near the edge of the survey area), which are small.
For reference,
the BOSS survey is expected to cover an area of about 30 times that of one
of our mocks, with about the same quasar density, so the errors for the
final BOSS results should be about $\sqrt{5/3}$ larger than in Figure 
\ref{fig:wedges}.


  As we can see in the left panel of Figure \ref{fig:wedges}, there is a 
disagreement
between the input theory and the measured correlation function that is
clearly above the errors. This disagreement is due to the bias of the
weights in equation \ref{eq:weight}: the weights are systematically smaller for
pixels with a smaller value of $\delta_F$.
 This causes the bias in the
estimator for the correlation function. To remove this bias, equation
\ref{eq:weight} can be modified to the following form, in which the 
weight is set equal to its value for $\delta_F=0$ and does
not depend on the measured flux:
\begin{equation}
 w'_i = \left\{ \sigma_F^2(z_i) + { A +
                  B(\lambda_i)\, \Delta\lambda \,
   \left[ \bar F(z_i)\, C(\lambda_i) + s(\lambda_i) \right] \over
   \left[ \bar{F}(z_i) C(\lambda_i) \right]^2 } \right\}^{-1} ~ .
\label{eq:unbiased}
\end{equation}
The results obtained using the weights $w'_i$ are shown in the right
panel of Figure \ref{fig:wedges}, showing a complete absence of any bias. This
example illustrates that particular care needs to be taken in the future
in the choice of weights to obtain unbiased estimates of the correlation
function: a better estimator than the one used in \cite{XiPush} will be 
required for a proper interpretation of the measurements with the final BOSS
data set.



%
%
 According
to this prediction, the amplitude of the BAO peak should be detectable
at the $\sim 5-\sigma$ level in bins of $10 h^{-1}\mpc$ in $r$
in the correlation function if all the data obtained in BOSS is as
expected, in agreement with previous authors.

\subsection{Variations in the Survey Strategy}

An application of the mock \lya forest surveys is to calculate the
precision achieved in the measurement of the correlation function on
large scales as a function of any survey properties in order to optimize
the design of the survey. This study may often be done using a Fisher
matrix approach without the need to generate survey realizations, but
using the mocks presented here allows one to include any possible
systematic effects in a more complete way.

\begin{figure}[h!]
 \begin{center}
  \includegraphics[scale=0.7, angle=-90]{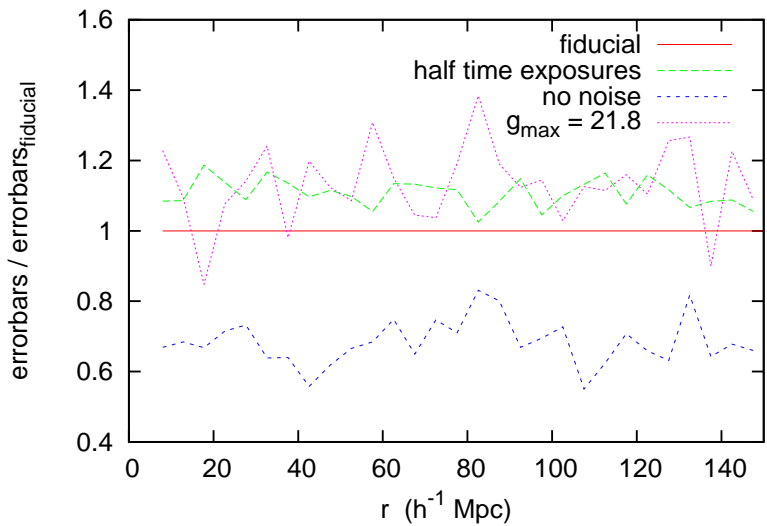}
 \end{center}
 \caption{Fractional change in the errorbars of the
 correlation function, for each radial bin, with respect to the fiducial
 survey, when varying survey parameters. The dashed green line assumes
 that only half as many exposures are obtained (i.e., the
 signal-to-noise is reduced by a factor $\sqrt{2}$), and the dotted pink
 line shows the result of eliminating the faintest 16\% of the quasars
 with $21.8 < g < 22$. The dotted blue line is for the case with no
 observational noise. }
 \label{fig:for}
\end{figure}

Here we study the change in the errorbars of the
correlation function when we vary either the exposure time or the number
of observed quasars within a fixed area. 

We note that the variation of
these errorbars with the area of the survey, if we keep the quasar
density fixed, is basically proportional to the inverse square root of
the area, apart from the presence of edge effects, which are already
small at the BAO scale for our fiducial survey with an area of
$300$ deg${}^2$.

 Figure \ref{fig:for} shows that the fiducial survey has errors that
are reduced by $\sim 30$ \% if the observational noise (both photon and
read-out noise in the detectors) were entirely eliminated. In other
words, the errors arising from observational noise and from the
intrinsic sampling variance in the \lya forest are comparable in our
fiducial survey. The best strategy to reduce the sampling variance is
to aim for the largest possible survey area. Increasing the source
density is more difficult because one has to search for fainter quasars,
which are harder to identify and have larger observational noise for a fixed 
exposure. The curves in Figure \ref{fig:for} show that reducing the number of
exposures by a factor of 2 degrades the error bars by the same amount (10
to 15\%) as eliminating the faintest 16\% of the quasars, in the
magnitude range $21.8 < g < 22$. Therefore, this shows that maximizing
the number of quasars that are observed is the best survey strategy,
even near the magnitude limit of the BOSS spectroscopic quasar survey
(see \cite{rossprep}), and even if this is done at the cost of some
reduction in the exposure time.

  A simple Fisher matrix approach was used in \cite{2007PhRvD..76f3009M}
to study the best survey strategy to measure the angular diameter
distance $D_A(z)$ and the Hubble parameter $H(z)$ from the BAO wiggles in
the power spectrum.
In their Figure 1, these authors show that when the survey limiting
magnitude is reduced from $g=22$ to $g=21.8$, the fractional error on the 
angular distance $D_A(z)$ increases by $\sim 20 \%$ and the Hubble parameter 
$H(z)$ increases by $\sim 10\%$, in agreement with the $10-15\%$ increase
of the errorbars that we find
(the S/N used for their figure is higher than in our mocks,
so their improvement for a fainter limiting magnitude should be
slightly higher than ours). 
In their Figure 5, \cite{2007PhRvD..76f3009M} show that the fractional
error on both scales increases by $\sim 10 \%$ if the $(S/N)^2$
is reduced by a factor of 2 (equivalent to reducing the number of exposures
by a factor 2), also in agreement with the $10-15\%$ increment of the
errorbars found in the analysis of our mocks. 
Similar results were obtained by \cite{2011arXiv1102.1752M}.
Our results therefore confirm these earlier studies, where we have now
included various effects in greater detail, such as the redshift
evolution, the expected length of the observable spectra and their
degree of overlap (these are typically included in Fisher matrix calculations
in a somewhat abstract, averaged way).

  Our method is highly flexible to allow for a rapid computation of the 
best strategy for survey optimization, including any systematic effects
that one may consider and include in the mocks in a realistic way. 
The mocks described here were already used to make a preliminary study
of the systematic effects of continuum fitting errors, spectroscopic
noise, metal absorption lines and high column density systems for the
first measurements with BOSS presented in \cite{XiPush}.

\subsection{Comparison with Gaussian field errorbars}

The forecasts for the accuracy of correlation function measurements with
the Fisher matrix approach in \cite{2007PhRvD..76f3009M} and
\cite{2011arXiv1102.1752M} assumed that the field can be modeled as a
Gaussian variable. We can test that this hypothesis does not indeed
alter the result by generating mocks of a Gaussian field and of a
field with the lognormal distribution of equation \ref{eq:lognormal}, with the
same power spectrum and generated with the same random numbers.

\begin{figure}[h!]
 \begin{center}
  \includegraphics[scale=0.7, angle=-90]{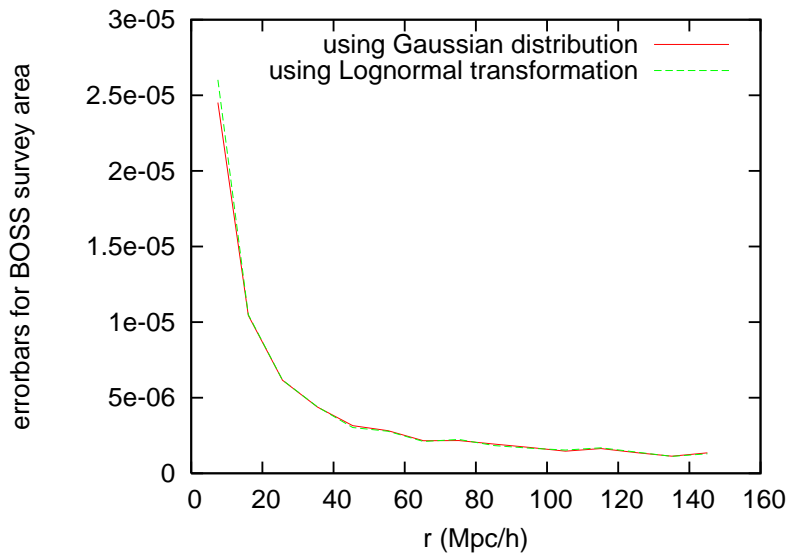}
 \end{center}
 \caption{Comparison of the errorbars in the angular averaged correlation 
 function expected for a survey similar to BOSS, compared to mocks where 
 the field has the same power spectrum but a Gaussian distribution.}
 \label{fig:vs}
\end{figure}

In Figure \ref{fig:vs} we show that the errorbars in the measurement of the
angular-averaged correlation function of a Gaussian mock survey are nearly
identical to the errorbars of a lognormal mock survey. This confirms the
expected result that the errorbars are not sensitive to the 1-point
function and are determined solely by the power spectrum (this does not, 
however, test the importance of longer range gravitational non-linearity
\cite{2009JCAP...08..020M}, although we do not expect this to be important on 
large scales either).

\section{Conclusions}

The method described here is able to create mock correlated spectra of
the \lya forest surveys mimicking the observed properties of real surveys.
Two free functions can be introduced as input to the mocks,
fixing the one-point distribution and two-point correlation function of
the field $\delta_F$, which can be made to evolve with redshift. The
higher order n-point functions that are not reproduced are not expected
to significantly affect the measurement errors of 2-point statistics on the 
large scales of interest.

  This paper presents only a simple example of the application of these
mocks to a survey with similar characteristics as BOSS. The technique
has already been used in the first analysis of BOSS data in 
\cite{XiPush}. In the future, we plan to improve our methodology to
apply it to a number of sources as large as the entire BOSS survey, and
to include all the observational effects in increasing detail. For example, 
one of the main applications of these mocks is to accurately model the effect
of high column density systems and metal-line absorption systems on the
measurement of the \lya forest correlation, which will be described in
\cite{LyaDLA}.

  Recently, \cite{2011MNRAS.tmp.1539G} and \cite{2011A&A...534A.135L} have
also presented methods for generating mock surveys of Lyman alpha absorption.
Both of these studies generate Gaussian fields in a 3D grid for the
density and the velocity field, and compute the transmitted flux using a
lognormal transformation. As mentioned in our Section 2, these methods
face computational challenges owing to the large amount of memory that
is required, which limits their study to small volumes (area of 79
${\rm deg}^2$ in \cite{2011MNRAS.tmp.1539G}) or poor resolution
(cell size of $3.2\, h^{-1}$ Mpc/h in \cite{2011A&A...534A.135L}).
Furthermore, they cannot easily produce large numbers of realizations
of a survey with the observed range of redshifts that are truly
independent of each other, because of the difficulty in running many
independent 3D simulations and storing their outputs. 




  An important difficulty of our method is still the memory needed
to store the covariance matrices of each k-mode, because their size
grows with the square of the number of simulated lines of sight. This
limits the number of lines of sight to $\sim 10^4$ when running the
code in a 24-CPU node with 32 Gb of shared memory (the CPU time used is 
$\sim 100$ hours). 
We are currently working on an improved version of the code that will be
able to avoid this limitation by using MPI and sparse matrix algorithms.



\begin{acknowledgments}
The simulations in this work were performed on CITA's Sunnyvale clusters 
which are funded by the Canada Foundation for Innovation, the Ontario 
Innovation Trust, and the Ontario Research Fund.
The authors thank Anze Slosar, Jean-Marc LeGoff and Nicolas Busca for
very helpful discussions.
This work was supported in part by Spanish grant AYA2009-09745.

\end{acknowledgments}

\bibliography{cosmo,cosmo_preprints,xip,font}
\bibliographystyle{JHEP}

\end{document}